\documentclass[journal]{IEEEtran}
\ifCLASSINFOpdf
\else
\fi
\usepackage{graphicx}

\usepackage{algorithm}
\usepackage[noend]{algpseudocode}
\usepackage{algorithmicx}
\usepackage{hyperref}

\usepackage{amsmath}
\usepackage{amssymb}
\begin{document}
%
\title{Probabilistic Fault-Tolerant Robust Traffic Grooming in OTN-over-DWDM Networks
}
%
%
%

\author{Dimitrios~Michael~Manias\textsuperscript{1}, Joe~Naoum-Sawaya\textsuperscript{2},  Abbas~Javadtalab\textsuperscript{3}, and~Abdallah~Shami\textsuperscript{1}\\ Western University\textsuperscript{1}, Ivey Business School\textsuperscript{2}, Concordia University\textsuperscript{3} \\
\{dmanias3, Abdallah.Shami\}@uwo.ca; jnaoum-sawaya@ivey.ca; abbas.javadtalab@concordia.ca
\thanks{© 2024 IEEE.  Personal use of this material is permitted.  Permission from IEEE must be obtained for all other uses, in any current or future media, including reprinting/republishing this material for advertising or promotional purposes, creating new collective works, for resale or redistribution to servers or lists, or reuse of any copyrighted component of this work in other works.}
}

%
%

\markboth{DRCN 2024}%
{Manias \MakeLowercase{\textit{et al.}}:Probabilistic Fault-Tolerant Robust Traffic Grooming in OTN-over-DWDM Networks}
%



\maketitle

\begin{abstract}
The development of next-generation networks is revolutionizing network operators' management and orchestration practices worldwide. The critical services supported by these networks require increasingly stringent performance requirements, especially when considering the aspect of network reliability. This increase in reliability, coupled with the mass generation and consumption of information stemming from the increasing complexity of the network and the integration of artificial intelligence agents, affects transport networks, which will be required to allow the feasibility of such services to materialize. To this end, traditional recovery schemes are inadequate to ensure the resilience requirements of next-generation critical services given the increasingly dynamic nature of the network. The work presented in this paper proposes a probabilistic and fault-tolerant robust traffic grooming model for OTN-over-DWDM networks. The model's parameterization gives network operators the ability to control the level of protection and reliability required to meet their quality of service and service level agreement guarantees. The results demonstrate that the robust solution can ensure fault tolerance even in the face of demand uncertainty without service disruptions and the need for reactive network maintenance. 
\end{abstract}


%
\IEEEpeerreviewmaketitle

\section{Introduction}
%
%
%
%
\IEEEPARstart{F}{ifth} Generation (5G) and beyond networks are characterized by improved performance requirements and diverse service and application capabilities. However, achieving such improvements over existing networking technologies requires stringent performance requirements, the profound integration of artificial intelligence in various levels of the network, and an array of enabling technologies, such as Multi-Access Edge Computing (MEC) \cite{giordani2020toward, tomkos2020toward}. In terms of performance requirements, next-generation networks require attributes such as low latency, high device density, and increasing levels of reliability. Furthermore, each successive networking generation introduces increasingly stricter limits for such network attributes as well as new Key Performance Indicators (KPIs) to monitor the efficiency of the network. \par
 
As artificial intelligence integration occurs, the timely and efficient communication of network-generated data is critical. As networks increase in size, complexity and dynamicity, the amount of network-generated data increases greatly \cite{9210141}; to this end, the rate at which network-generated data is produced and consumed rapidly increases. This increase threatens artificial intelligence integration as effective machine learning models need to be built using a comprehensive dataset representative of the network while also conducting inference on the most recently generated data. Already, before the mass deployment of 5G networks, one of the critical KPIs that is emerging for 6G networks is called the Age of Information (AoI) \cite{AOI1, AOI2, AOI3}. The AoI defines the time between data generation and data consumption. As such, ensuring any intelligence model or agent receives the freshest information in a timely manner to make predictions and recommendations promptly is critical. This is where transport networks, specifically Optical Transport Networks (OTN) leveraging Dense Wavelength Division Multiplexing (DWDM), are required to reduce the AoI metric and handle the continually increasing network demand \cite{manias2023role}. \par


In OTN service management, resilience schemes are generally defined by the primary service path, the existence of a secondary service path, as well as the existence of a reactive recovery plan. For example, a 1 + 0 recovery scheme indicates that there exists a primary path with no recovery path or recovery plan present. A 1 + R recovery scheme indicates that there exists a primary path with a reactive recovery plan that will reconfigure network traffic in the event of a fault or failure. Finally, a 1 + 1 recovery plan means that there exists both a primary and backup path with dedicated reserved capacity. In practice, network operators tend to gravitate towards a 1 + R recovery scheme stating that 1 + 1 recovery is wasteful as excessive reserve capacity reduces optical channel utilization and is costly. Given the nature of next-generation critical services, a 1 + R recovery scheme is inadequate as it does not ensure service continuity. \par

As mentioned, next-generation networks are expected to be more dynamic than previous networking generations. This dynamicity, coupled with the development and emergence of new use cases, services, and applications, leads to great uncertainty in the demand. When performing MANO activities, it is naïve to assume a static traffic matrix and a snapshot of current network conditions is representative of the behaviour of the network, as done in the past. To this end, techniques such as robust optimization must be leveraged to protect the derived solutions and subsequent MANO recommendations against demand uncertainty. \par

The contributions of the presented work can be summarized as follows:

\begin{itemize}
\item The development of a flexible traffic grooming and infrastructure model capable of addressing the reliability requirements and fault tolerance of next-generation critical services.
\item The use of robust optimization methods to address the demand uncertainty challenge facing network operators.
\item The complete parameterization of the developed model giving the network operator flexibility to control the level of overprovisioning for reserve capacity and demand uncertainty. 
\end{itemize}

The remainder of this paper is structured as follows. Section II presents the state-of-the-art regarding traffic grooming, infrastructure placement and network resilience schemes for OTN-over-DWDM networks. Section III presents background information and discusses the system model. Section IV presents the robust optimization model formulation. Section V discusses the experiment setup and the parameters used. Section VI presents and analyzes the results of the experiment. Finally, Section VII concludes the paper and discusses avenues for future work.

\section{Related Work}
In terms of traffic grooming and infrastructure placement, the problems are structured as optimization models with solutions achieved through linear programming methods, meta-heuristic optimization methods, computational intelligence, and near-optimal heuristic solutions. The objectives of these optimization models generally consider meeting a set of demands while minimizing a set of deployed network equipment and infrastructure. Attarpour \textit{et al.} \cite{attarpour2022minimizing} consider the cost minimization of hierarchical traffic grooming boards deployed in mesh networks. Moniz \textit{et al.} \cite{moniz2019multi} explore the cost-effective placement of OTN switches through OTN interface minimization. Ramachandran \textit{et al.} \cite{ramachandran2019capacity} discuss the joint optimization objective of minimizing the total number of cross-connections, while simultaneously maximizing the utilization of aggregated and groomed traffic on network channels. Ibrahimi \textit{et al.} \cite{ibrahimi2022minimizing} attempt to minimize the equipment and energy costs in filterless metro networks through the minimization of the required logical node infrastructure and transponders. \par
The following works have explored the topics of restoration and protection in multi-layered networks. Xavier \textit{et al.} \cite{xavier2021heuristic} explore the use of a heuristic algorithm for sharing restoration interfaces in OTN-over-DWDM networks. Tan \textit{et al.} \cite{tan2022designing} consider the design of OTN-over-DWDM networks with multi-failure recovery, while considering both the electrical and optical domains. Oliveira \textit{et al.} \cite{oliveira2015comparison} compare network protection schemes in multi-layer networks. The authors use a cost model to determine the cost of protecting each layer independently in the considered network architecture. Le \textit{et al.} \cite{le2020survivable} explore survivability in optical metro networks through virtual network over physical network mapping and content replication. The authors use content connectivity, a metric that assesses the reach of a specific piece of content from every node in a network, instead of the traditionally used network connectivity, which is the reachability of a node from all other nodes. Furthermore, Arpanaei \textit{et al.} \cite{arpanaei2022comparative} perform a comparative study on shared backup and restoration paths in elastic optical networks. This study aims to minimize a composite cost function related to the number of transceivers and frequency slots used to establish a lightpath. \par

The state-of-the-art for OTN-over-DWDM traffic grooming, infrastructure placement, and resiliency schemes is dominated by deterministic solutions. To this end, the proposed solutions are prone to demand uncertainty caused by the dynamicity of next-generation networks and applications. Demand uncertainty is a challenge facing network operators as the initial demand planned for is different from the realized demand. This can have crippling effects on network operations as an increase in the required demand can require reactive network maintenance and can lead to blocked requests and even outages. Furthermore, given the reduced time to market for new use cases and applications, the nature of the user’s behaviour regarding their demand habits is prone to change. These phenomena, coupled with activities such as crowd events, can cause various demand changes, some instantaneous and short-lived, others long-term and unwavering.

Regarding resilience, many proposed schemes consider 1+R resilience while claiming that 1+1 resilience schemes are too costly. Given the variety of critical services expected in 5G and beyond networks with their extremely stringent performance and reliability guarantees coupled with the detrimental impact service interruption can have, 1+1 is the minimum acceptable level of system resilience for such services. To this end, beyond 1+1 resilience is seldom considered and, even when considered, does not provide flexibility to the network operator.

The work presented in this paper aims to address the aforementioned gaps and unaddressed challenges in the state-of-the-art. This work proposes a flexible, robust, probabilistic, and fault-tolerant traffic grooming and infrastructure placement solution for OTN-over-DWDM networks. The challenge of demand uncertainty is directly addressed through robust optimization, a technique used in operations research to protect solution feasibility within specified levels of demand uncertainty. To this end, the demand matrix considered during the infrastructure planning and traffic grooming is not considered as being static, and the worst-case deviations that can negatively affect the solution are anticipated and protected against. The proposed model enables the network operator to select a number of independent backup paths that are exclusively used to ensure the reliability of a specific demand. Finally, a probabilistic coefficient is used to control the level of overprovisioning through reserved capacity in the system. By considering these aspects in the proposed solution, network operators can create a resilient and robust infrastructure planning and traffic grooming solution that can remain feasible in the face of demand uncertainty and network fault scenarios.

\section{System Model}
\begin{figure}[!htbp]
\centerline{\includegraphics[width=\columnwidth]{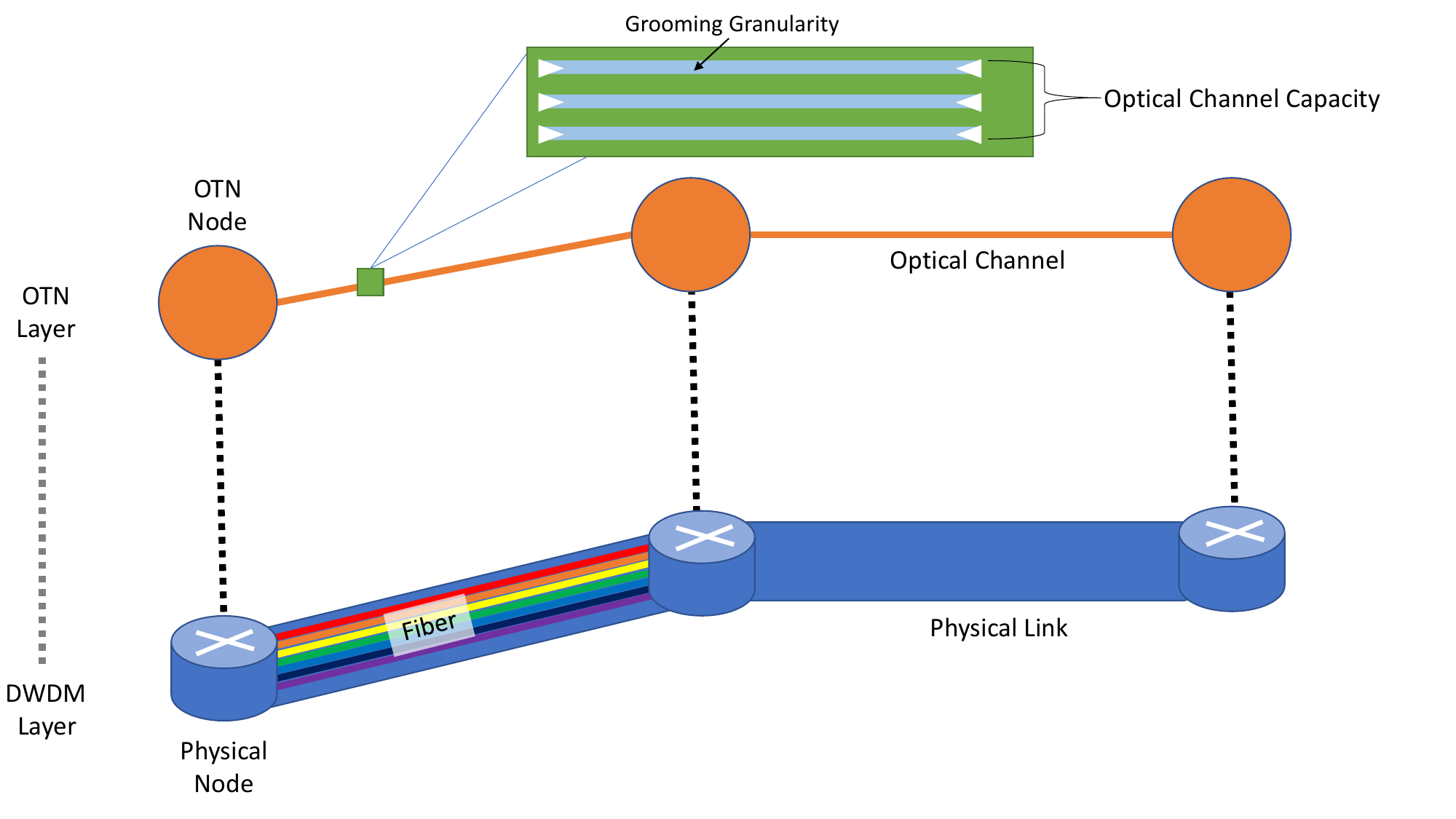}}
\caption{OTN-over-DWDM System Model}
\label{model}
\end{figure}

This section outlines the system model and some background information pertinent to the presented work. Regarding the system model, Fig. \ref{model} outlines the basic concepts. As seen in this figure, an OTN-over-DWDM architecture is presented. The DWDM layer comprises physical nodes connected by physical links with fiber cables. The OTN layer is overlayed on top of the physical layer and comprises OTN nodes connected through optical channels. Two line interface cards are required to create an optical channel, one at the source node and one at the destination node. An optical channel is associated with a wavelength in the DWDM layer. A combination of optical channels using different wavelengths can be used to complete a service request by deploying an OTN switch which enables wavelength conversion at intermediate nodes. In terms of the optical channel itself, its capacity is determined by the modulation format it can support, which depends on the physical distance between its source and destination nodes. For traffic grooming, the capacity of an optical channel is often expressed as the number of granularity slots it can support \cite{moniz2019multi}. By having smaller granularity, a finer level of grooming can be achieved. As seen in Fig. \ref{model}, the capacity of the optical channel is the summation of all the grooming granularity slots.

One of the objectives of this work is to ensure reliability in the case of a fault or failure to protect critical services against downtimes and interruptions. As mentioned, the disruption of a critical service (\textit{i.e.,} financial, emergency, transportation) can have devastating impacts on the public, in terms of safety, as well as the network operator in the form of severe financial repercussions and liability. Rak \textit{et al.} classified the major external events that can lead to massive optical network failures \cite{rak2021disaster}. Events, including natural disasters and disruptions caused by inclement weather, can severely impact the network's operation. Furthermore, the risk of malicious attacks aimed at disrupting service and taking down the network is more dangerous than ever. To this end, traditional optical network resilience schemes may prove incapable of maintaining service continuity in the face of such events. \par



The proposed work considers multiple independent backup paths with only a common source and destination node. This path independence can prevent single or multi-failure scenarios depending on the number of backup paths.By considering multiple independent backup paths, increasing levels of service resilience are introduced to protect against multiple node/link failure scenarios. \par


The proposed work also considers giving the network operator the flexibility to control the level of overprovisioning in the system related to reserve capacity. In this work, overprovisioning stems from the robust formulation related to the demand uncertainty and the reserve capacity from the backup path(s). The amount of overprovisioning related to the demand uncertainty is directly related to the robust parameters $\Gamma$ and $\hat{B_d}$, which determine the number of demands allowed to deviate in the solution and the amount by which they deviate off of their nominal value, respectively. The nominal value of a demand in this case is its requested/planned value. Demand uncertainty causes a deviation in this value and therefore, a change is observed. 

In the proposed formulation, $\mu_r$ is introduced as the resilience coefficient for the backup service path and controls the overall level of overprovisioning in the system (both for robust and backup reserve capacity). 
Setting $\mu_r$ to a value greater than one increases the overall level of overprovisioning in the system as both the robust and backup capacities have increased. Conversely, setting this parameter to a value less than one reduces the level of overprovisioning as the robust and backup capacities are smaller. It should be noted that the $\mu_r$ parameter counteracts the robust parameters in this formulation. If this is unfavourable, the $\mu_r$ can be removed from the robust constraints (expressed in the formulation in the following section), and it will only affect the overprovisioning related to the backup reserve capacity. The reason for its inclusion in robust capacity is to allow the operator to scale the level of overprovisioning as a whole. \par

\section{Robust Optimization Model Formulation}
This section outlines the proposed optimization model. The objective of this model is to perform infrastructure placement and traffic grooming based on a set of demands while minimizing the number of optical channels created and OTN switches deployed. This model leverages robust optimization to address demand uncertainty and uses dedicated and independent backup paths to meet the resilience requirements of next-generation critical services. Additionally, a probabilistic coefficient is introduced to control the level of overprovisioning stemming from the dedicated backup path and the consideration of demand uncertainty. 

The various symbols and notations used in the robust optimization formulation are presented in Table \ref{notations}. In terms of the robust formulation, the objective function, as defined through Eq. \ref{rob_obj}, minimizes the weighted sum of optical channels created and OTN switches deployed. The coefficients $\alpha$ and $\beta$ are determined based on network operator requirements and priorities (\textit{i.e.,} cost). Equations \ref{rob_partial_flow} and \ref{rob_partial_flow_backup} ensure that the partial flow variables for the primary and backup paths can assume a value between zero and one. Effectively, these variables denote the fraction of demand $d$ that is groomed along the optical channel with source $i$ and destination $j$. In the case of multiple backup paths, $r$ distinguishes which path the variable corresponds to. Equation \ref{rob_optical_cap} is the optical channel capacity constraint and ensures that the sum of fractional flows using an optical channel across all paths plus the demand uncertainty is less than or equal to the capacity of the optical channel. In this equation, the $\Gamma$, $z$, and $p$ variables are used to express demand uncertainty. $\mu_r$ is the coefficient of resilience and is correlated to the probability of failure and the required level of system overprovisioning. In order to have a full backup (1 + 1) scheme, the value of this coefficient for a backup path must be 1. Equations \ref{rob_flow_cons} and \ref{rob_flow_cons_xi} are the flow conservation constraints and ensure that for all primary and backup paths, the entirety of the demand flows from source to destination. Equations \ref{rob_switchability} and \ref{rob_switchability_xi} determine if an OTN switch is required at node $i$. By order of superseding constraints, if any primary or backup path uses node $i$ as an intermediate node between two optical channels, an OTN switch is required, and $y_i$ assumes a value of 1. Equation \ref{rob_dwdm_cap} is the DWDM capacity constraint and ensures enough wavelengths in the DWDM layer to support the creation of optical channels in the OTN layer. Equations \ref{source_const} and \ref{dest_const} are the source and destination path exclusivity constraints and ensure that the primary and backup paths do not use the same optical channel when exiting the source node or entering the destination node. Equation \ref{interm_node_const} is the intermediate node path exclusivity constraint and ensures that all primary and backup paths use unique intermediate nodes without duplication. The combination of the path exclusivity constraints ensures that aside from a source/destination fault, a feasible backup path exists in the case of a single or multi-node fault along the intermediate path nodes. Equations \ref{binary_const_1} - \ref{binary_const_4} are binary variable constraints that ensure the variables $X$ and $\Xi$ assume a value of one if $x$ or $\xi$ are used for a given path, respectively. Equation \ref{rob_c_1} is used to determine the value of the robust variables used to represent demand uncertainty in the optical channel capacity constraint. Finally, Eq. \ref{rob_c_2} through \ref{rob_cont_var} are general variable constraints.

\begin{table}

\caption{List of Symbols and Notations \label{notations}}
\begin{tabular}{|p{1cm}|p{7cm}|}
\hline
\textbf{Symbol} & \textbf{Description}                                                                         \\ \hline
       $g$ & Granularity                                                                         \\ \hline
             $\alpha$ & Optical channel weight coefficient \\ \hline
      $\beta$ & Optical switch weight coefficient \\ \hline
       $\epsilon$ & Very small positive non-zero number \\ \hline
       $M$ & Very large positive non-zero number \\ \hline
         $\Gamma$ & Robust formulation solution conservativeness parameter                              \\ \hline

         $D$ & Set of demands                                                                      \\ \hline
       $O$ & Set of optical channels                                                             \\ \hline
       $R$ & Set of backup service paths                                                         \\ \hline
         $E$ & Set of edges in physical network                                                    \\ \hline
       $N$ & Set of nodes in physical network                                                    \\ \hline
           $\Phi_i$& Set of nodes forming optical channels with node $i$ as the source node                                              \\ \hline
       $\Psi_i$ & Set of nodes forming optical channels with node $i$ as the destination node                                         \\ \hline
       
            $\Delta_e$ & Set of optical channels that use edge $e$                                             \\ \hline

       $w_e$ & Number of wavelengths available on link $e$                                           \\ \hline
      $B_d$ & Bandwidth requirement for demand $d$                                                 \\ \hline

      $\mu_r$ & Resilience coefficient for backup service path $r$ \\ \hline

      $C_{ij}$ & Capacity of optical channel with source $i$ and destination $j$                         \\ \hline

             $src(d)$ & Source node for demand $d$ \\ \hline
       $dst(d)$ & Destination node for demand $d$ \\ \hline

       $\hat{B_d}$ & Deviation of $B_d$                                                               \\ \hline
      
      $x_{ij}^d$ & Fractional flow for demand $d$ across optical channel with source $i$ and destination $j$ used in the primary service path \\ \hline
      $\xi_{ij}^{dr}$ & Fractional flow for demand $d$ across optical channel with source $i$ and destination $j$ used in backup service path $r$\\ \hline
      $X_{ij}^d$ & Binary variable denoting if demand $d$ utilizes optical channel with source $i$ and destination $j$ in its primary service path \\ \hline
      $\Xi_{ij}^{dr}$ & Binary variable denoting if demand $d$ utilizes optical channel with source $i$ and destination $j$ in backup service path $r$\\ \hline
      $\theta_{ij}$ & Number of optical channels with source $i$ and destination $j$                         \\ \hline
      $y_n$ & Binary variable denoting if an OTN switch exists at node $n$ \\ \hline

       $p_{ij}^d$ & Robust dual variable                                                             \\ \hline
       $z_{ij}$& Robust dual variable     \\
       \hline

\end{tabular}
\end{table}

\textit{Objective:}

\begin{equation}
\label{rob_obj}
    minimize(\alpha \cdot \sum_{(i,j) \in O}\theta_{ij} + \beta \cdot \sum_{n \in N}y_n)
\end{equation}
\textit{Subject to:}

\begin{equation}
\label{rob_partial_flow}
    0 \leq x_{ij}^d \leq 1 \, \forall \, (i,j) \in O \,;\, d \in D
\end{equation}

\begin{equation}
\label{rob_partial_flow_backup}
    0 \leq \xi_{ij}^{dr} \leq 1 \, \forall \, (i,j) \in O \,;\, d \in D
\end{equation}

\begin{equation}
\label{rob_optical_cap}
\begin{split}
    \Gamma \cdot z_{ij} + \sum_{d \in D}{B_d \cdot (x_{ij}^d + \sum_{r \in R}{\mu_r \cdot \xi_{ij}^{dr}})} \\
    + \sum_{d \in D}{p_{ij}^d} \leq C_{ij} \cdot g \cdot \theta_{ij} \, \forall \, (i,j) \in O
    \end{split}
\end{equation}

\begin{equation}
\label{rob_flow_cons}
\begin{split}
    \sum_{j \in \Psi_i}{x_{ji}^d} - \sum_{j \in \Phi_i}{x_{ij}^d} = \begin{cases}
    -1 \quad & \text{if} \, i = src(d) \\
    1 \quad & \text{if}  \, i = dst(d) \\
    0 \quad & otherwise\\
    \end{cases} \\ \forall d \in D; \, i \in N
    \end{split}
\end{equation}

\begin{equation}
\label{rob_flow_cons_xi}
\begin{split}
    \sum_{j \in \Psi_i}{\xi_{ji}^{dr}} - \sum_{j \in \Phi_i}{\xi_{ij}^{dr}} = \begin{cases}
    -1 \quad & \text{if} \, i = src(d) \\
    1 \quad & \text{if}  \, i = dst(d) \\
    0 \quad & otherwise\\
    \end{cases} \\ \forall d \in D; \, i \in N; \, r \in R
    \end{split}
\end{equation}

\begin{equation}
    \label{rob_switchability}
    \begin{split}
    y_i \geq -1 \cdot |\sum_{j \in \Psi_i} {x_{ji}^d} - \sum_{j \in \Phi_i}{x_{ij}^d}| + \\ \epsilon \cdot \sum_{j \in \Psi_i}{x_{ji}^d} + \epsilon \cdot \sum_{j \in \Phi_i}{x_{ij}^d} 
    \, \forall \, i \in N \,; d \in D
    \end{split}
\end{equation}

\begin{equation}
    \label{rob_switchability_xi}
    \begin{split}
    y_i \geq -1 \cdot |\sum_{j \in \Psi_i} {\xi_{ji}^{dr}} - \sum_{j \in \Phi_i}{\xi_{ij}^{dr}}| + \\ \epsilon \cdot \sum_{j \in \Psi_i}{\xi_{ji}^{dr}} + \epsilon \cdot \sum_{j \in \Phi_i}{\xi_{ij}^{dr}} 
    \, \forall \, i \in N \,;  r \in R \,;d \in D
    \end{split}
\end{equation}

\begin{equation}
\label{rob_dwdm_cap}
    \sum_{(i,j) \in \Delta_e}{\theta_{ij}} \leq w_e \, \forall \, e \in E
\end{equation}

\begin{equation}
\label{source_const}
\begin{split}
    \sum_{j \in \Phi_{src(d)}}{X_{src(d)j}^d} + \sum_{j \in \Phi_{src(d)}}{\sum_{r \in R}{\Xi_{src(d)j}^{dr}}} = \| R \| + 1 \\ \forall \,  d \in D
    \end{split}
\end{equation}

\begin{equation}
\label{dest_const}
\begin{split}
    \sum_{j \in \Psi_{dst(d)}}{X_{jdst(d)}^d} + \sum_{j \in \Psi_{dst(d)}}{\sum_{r \in R}{\Xi_{jdst(d)}^{dr}}} = \| R \| + 1 \\ \forall \,  d \in D
    \end{split}
\end{equation}

\begin{equation}
\label{interm_node_const}
\begin{split}
    \sum_{j \in N \backslash \{src(d), dst(d)\}}{X_{ij}^d + X_{ji}^d + \sum_{r \in R}{\Xi_{ij}^{dr} + \Xi_{ji}^{dr}}} \leq 1  \\ \forall \,  d \in D; \, i \in N \backslash \{src(d), dst(d)\} 
    \end{split}
\end{equation}

\begin{equation}
\label{binary_const_1}
    x_{ij}^d \cdot M \geq X_{ij}^d \, \forall \, (i,j) \in O \,;\, d \in D
\end{equation}

\begin{equation}
\label{binary_const_2}
    \xi_{ij}^{dr} \cdot M \geq \Xi_{ij}^{dr} \, \forall \, (i,j) \in O \,;\, d \in D; \, r \in R
\end{equation}

\begin{equation}
\label{binary_const_3}
    x_{ij}^d \leq X_{ij}^d \, \forall \, (i,j) \in O \,;\, d \in D
\end{equation}

\begin{equation}
\label{binary_const_4}
    \xi_{ij}^{dr} \leq \Xi_{ij}^{dr} \, \forall \, (i,j) \in O \,;\, d \in D; \, r \in R
\end{equation}

\begin{equation}
\label{rob_c_1}
\begin{split}
    p_{ij}^d + z_{ij} \geq \hat{B_d} \cdot (x_{ij}^d + \sum_{r \in R}{\mu_r \cdot \xi_{ij}^{dr}}) \\ \forall \, (i,j) \in O \,;\, d \in D
    \end{split}
\end{equation}

\begin{equation}
\label{rob_c_2}
    p_{ij}^d \geq 0 \, \forall \, (i,j) \in O \,;\, d \in D
\end{equation}

\begin{equation}
\label{rob_c_3}
    z_{ij} \geq 0 \, \forall \, (i,j) \in O
\end{equation}

\begin{equation}
\label{rob_int_var}
    \theta_{ij} \in \mathbb{Z}_0^+
\end{equation}

\begin{equation}
\label{rob_y_var}
    y_n, \, X_{ij}^{d}, \, \Xi_{ij}^{dr} \, \in \{0,1\}
\end{equation}

\begin{equation}
\label{rob_cont_var}
    x_{ij}^d, \, \xi_{ij}^{dr}, \, p_{ij}^d, \, z_{ij} \in \mathbb{R}_0^+
\end{equation}

\section{Experiment Setup}
\begin{figure}[!htbp]
\centerline{\includegraphics[width=0.8\columnwidth]{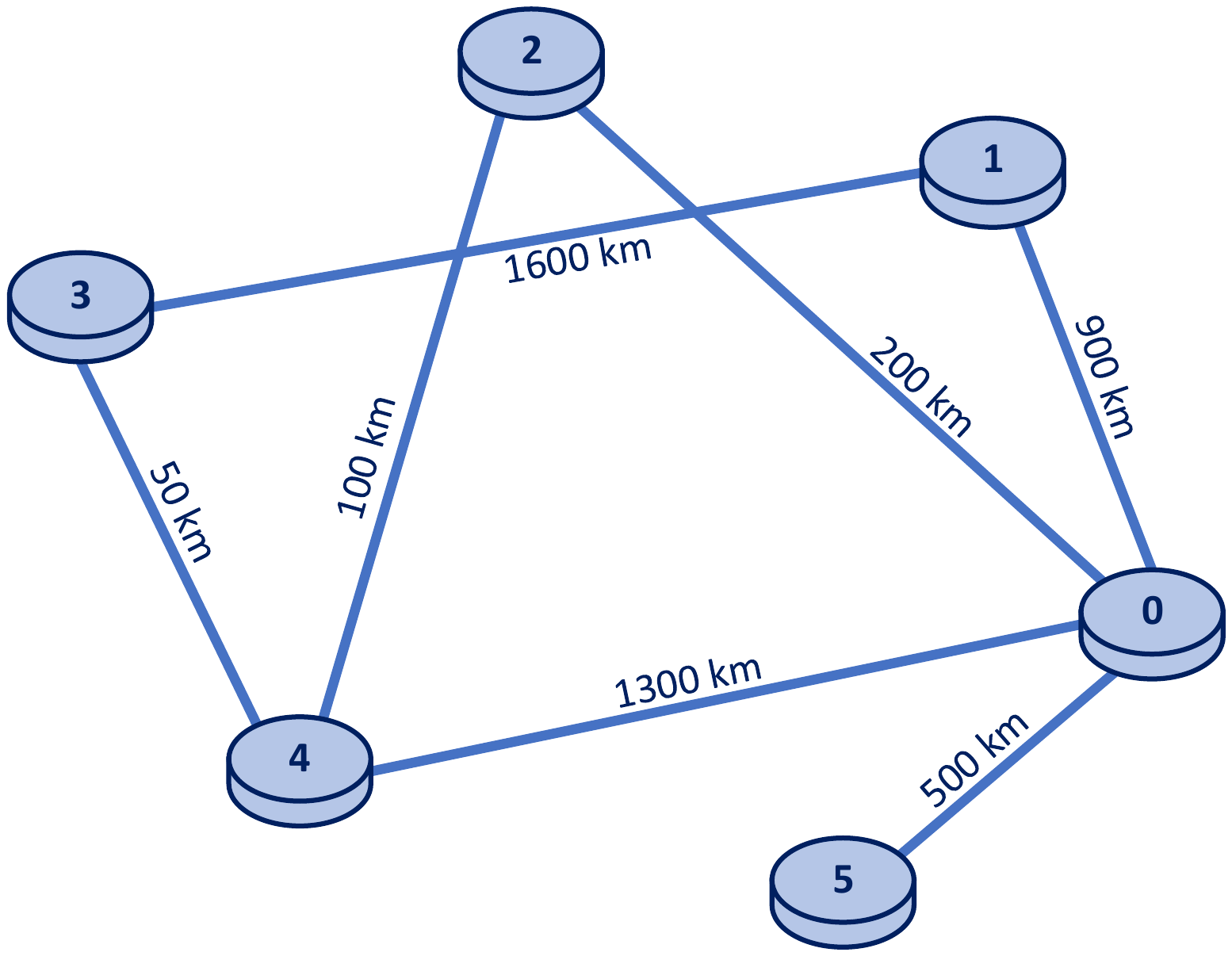}}
\caption{6-Node Network Topology}
\label{6N}
\end{figure}

This section outlines the experiment that was conducted, for which results are presented in the following section. It should be noted that the proposed optimization model is applicable to any network topology with any parameter settings. For the experiment, both the 6-node (6N) network topology depicted in Fig. \ref{6N} as well as the common NSFNET topology were used. The results were consistent across both topologies, however the presented results pertain to the 6N topology, which was selected for its ease of presentation and to demonstrate the fundamental mechanics of the proposed optimization model. 

A demand matrix of 50 demands was used, with the required demand generated uniformly from 10 Gb/s to 50 Gb/s. The grooming granularity was selected to be 1.25 Gb/s. The number of wavelengths on a given physical link is 80. Given the network size, the objective coefficients $\alpha$ and $\beta$ are selected to be 1 and 0, respectively, as it is expected that every node will need an OTN switch. For the sake of completeness, the readers are referred to previous work by the authors with an extensive analysis of the effect of the objective coefficients on the resulting grooming solution \cite{review} regarding the primary service path. The capacity of each optical channel is derived from the physical distance between nodes and the modulation format it can support, concurrent with industry standards \cite{7740837}. For this work, the robust solution protects against 20\% of the demands deviating by 10\% off their nominal value. The readers are referred to our previous work for further analysis into varying a larger number of demands with more significant deviation \cite{review}. The experiment presented in this work was conducted with 1 Primary Path + 1 Backup Path ($\mu_1 = 1$).

\section{Results and Analysis}

\subsection{Critical Node Analysis}
\begin{figure}[!htbp]
\centerline{\includegraphics[width=0.85\columnwidth]{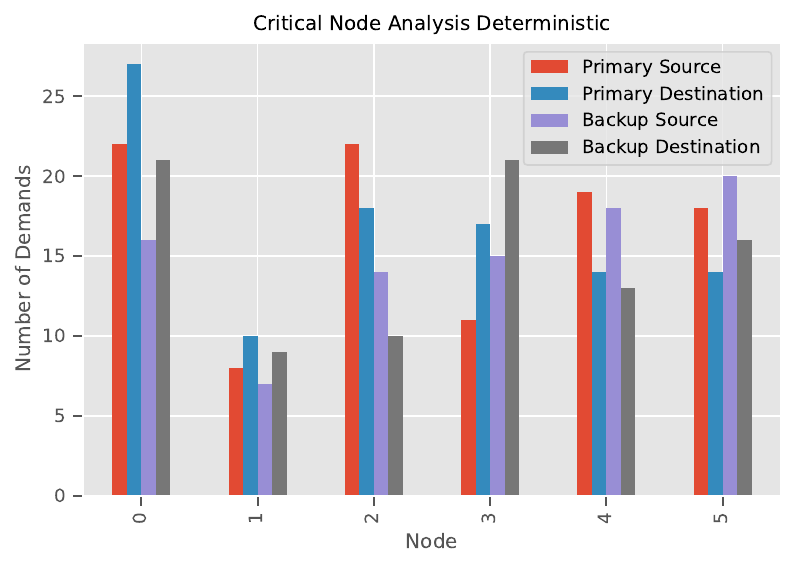}}
\caption{Critical Node Analysis (Deterministic)}
\label{crit_node_deterministic}
\end{figure}

\begin{figure}[!htbp]
\centerline{\includegraphics[width=0.85\columnwidth]{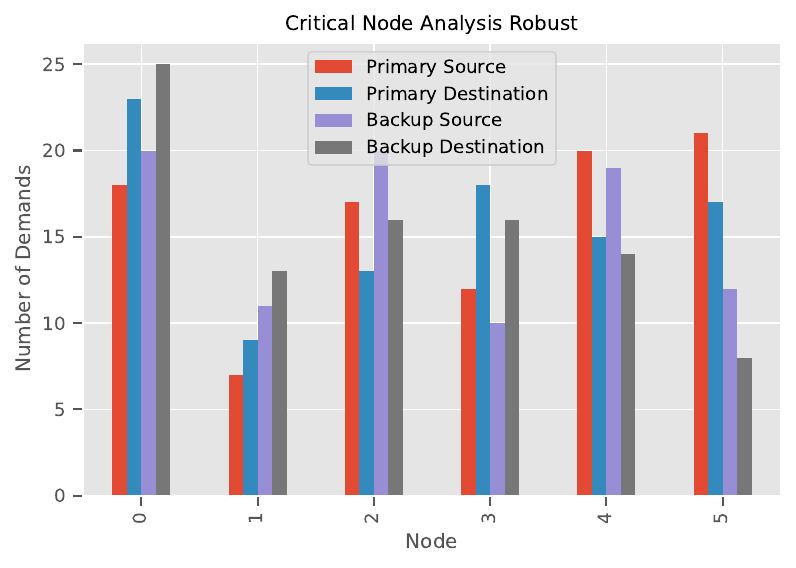}}
\caption{Critical Node Analysis (Robust)}
\label{crit_node_robust}
\end{figure}

The results presented in Fig. \ref{crit_node_deterministic} and Fig. \ref{crit_node_robust} consider a high-level overview of the system as they present a critical node analysis. Figure \ref{crit_node_deterministic} presents the critical node analysis results for the deterministic solution, and Fig. \ref{crit_node_robust} presents the critical node analysis for the robust solution. In these figures, the criticality of a node is determined based on the number of demands that pass through each node along the primary and backup paths. As seen in these figures, node 1 is deemed the most critical in both grooming scenarios since more demands use it to complete their service. The criticality of the remaining nodes varies slightly among the two solutions, further reinforcing the statement that the level of robustness affects the resulting grooming solution. Understanding the criticality of each node through such an analysis is critical to understanding the system's behaviour as a whole. Measures can be taken proactively to ensure the most critical nodes have the greatest level of resilience, as their fault would have a more significant impact on network operations compared to less critical nodes.

\subsection{Fault Analysis}
\begin{figure}[!htbp]
\centerline{\includegraphics[width=0.85\columnwidth]{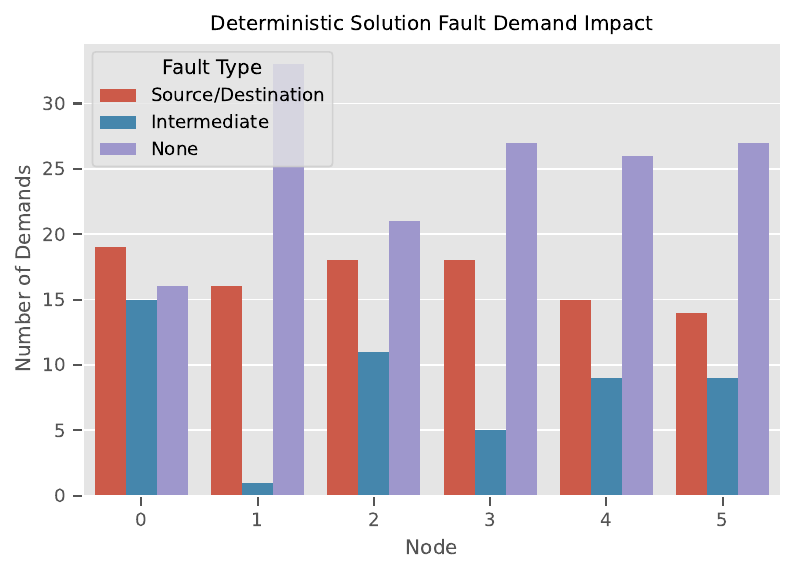}}
\caption{Fault Effect Analysis (Deterministic)}
\label{deterministic_fault_effect}
\end{figure}

\begin{figure}[!htbp]
\centerline{\includegraphics[width=0.85\columnwidth]{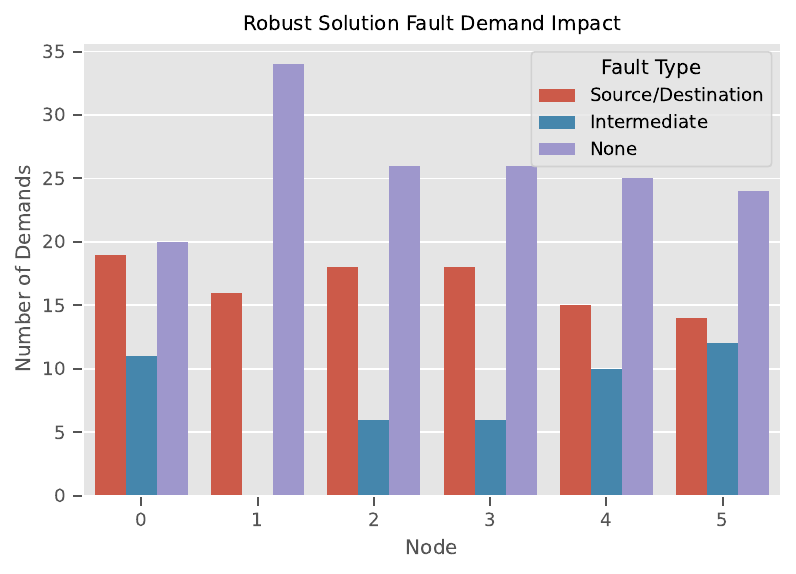}}
\caption{Fault Effect Analysis (Robust)}
\label{robust_fault_effect}
\end{figure}

The next set of results is presented in Fig. \ref{deterministic_fault_effect} and Fig. \ref{robust_fault_effect}. Through these figures, a fault analysis is conducted on a per-node basis. This analysis considers the impact of a fault occurring on a given node. As previously mentioned, there are three possible scenarios, a source/destination fault, an intermediate fault, or no fault. In the case of a source/destination fault, the service cannot be completed since its demand cannot leave the source node or reach the destination node. In the case of an intermediate fault on the primary path, the backup path is used due to the path exclusivity constraints. Finally, if no fault has occurred, the primary path is used. This analysis is important as it helps further the understanding of the system behaviour by determining the number of demands that will not be met due to a source/destination fault and the number of demands that need to be met using their backup path due to an intermediate fault.

\begin{figure*}[!htbp]
\centerline{\includegraphics[width=1.85\columnwidth]{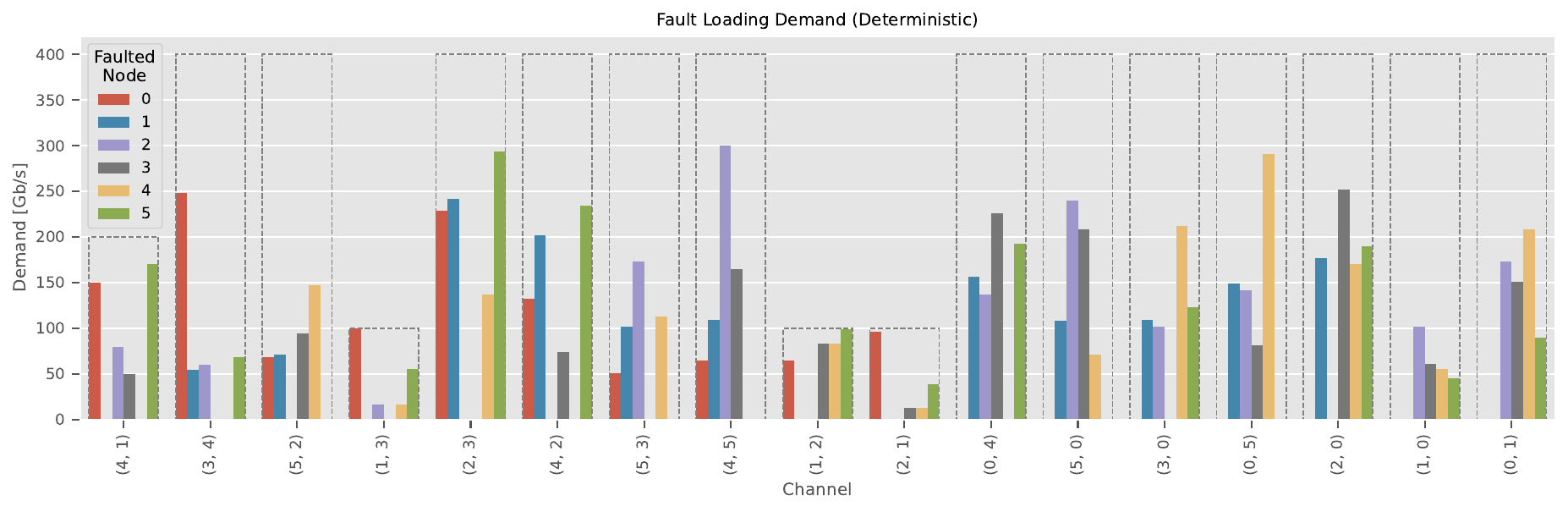}}
\caption{Post-Fault Loading (Deterministic)}
\label{deterministic_load}
\end{figure*}

\begin{figure*}[!htbp]
\centerline{\includegraphics[width=1.85\columnwidth]{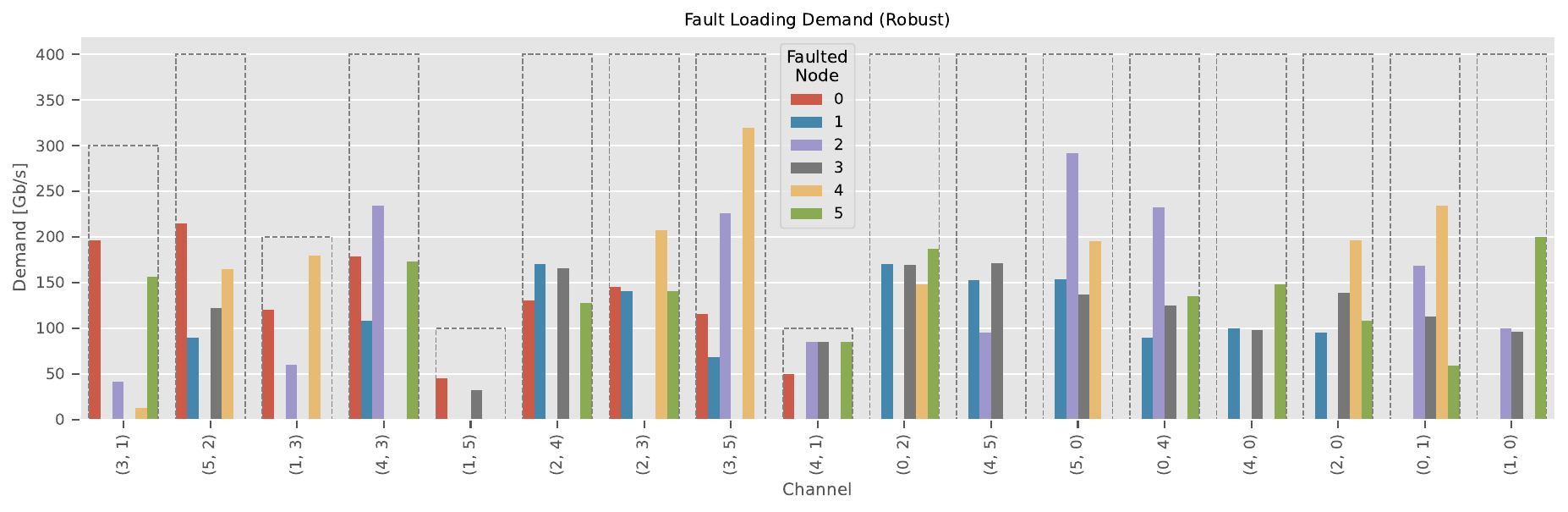}}
\caption{Post-Fault Loading (Robust)}
\label{robust_load}
\end{figure*}

A logical extension of this analysis is to compare the post-fault loading for each optical channel. To do this, a single node fault is simulated, and the demands that have not experienced a source/destination fault are groomed based on the optimization solution. When a demand has experienced an intermediate node fault on its primary path, its backup path is used. Figure \ref{deterministic_load} presents the post-fault loading analysis for the deterministic solution, and Fig. \ref{robust_load} presents the post-fault loading analysis for the robust solution. In each figure, the x-axis denotes the optical channel, and the y-axis denotes the demand in Gb/s. The dotted bar values denote the capacity of the optical channel. Each single-node fault is denoted by the colours as described in the legend. A critical distinction between the robust and deterministic solutions is that, in general, the deterministic solution tends to approach the capacity of the optical channel. In contrast, the robust solution tends to leave room to account for the possible demand deviation due to uncertainty. This observation is evident in the deterministic solution in optical channels (1,3), (1,2), and (2,1) compared to the optical channel (4,1) in the robust solution. Exploring the post-fault loading is a critical step toward understanding the system's behaviour during a fault. Furthermore, the difference in the percentage of maximum optical channel loading between the robust and deterministic solutions leads to an analysis of how demand uncertainty will impact the post-fault performance of the system.

\subsection{Deviation Analysis}
\begin{figure*}[!htbp]
\centerline{\includegraphics[width=1.85\columnwidth]{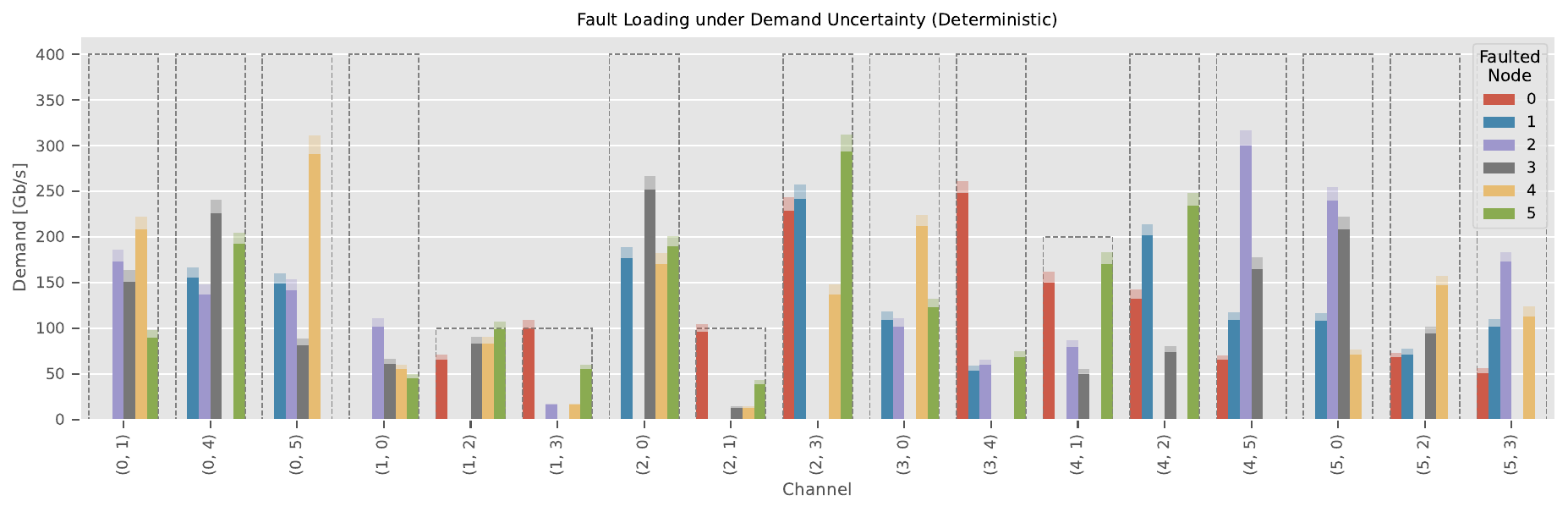}}
\caption{Post-Fault Deviation (Deterministic)}
\label{deterministic_dev_post}
\end{figure*}

\begin{figure*}[!htbp]
\centerline{\includegraphics[width=1.85\columnwidth]{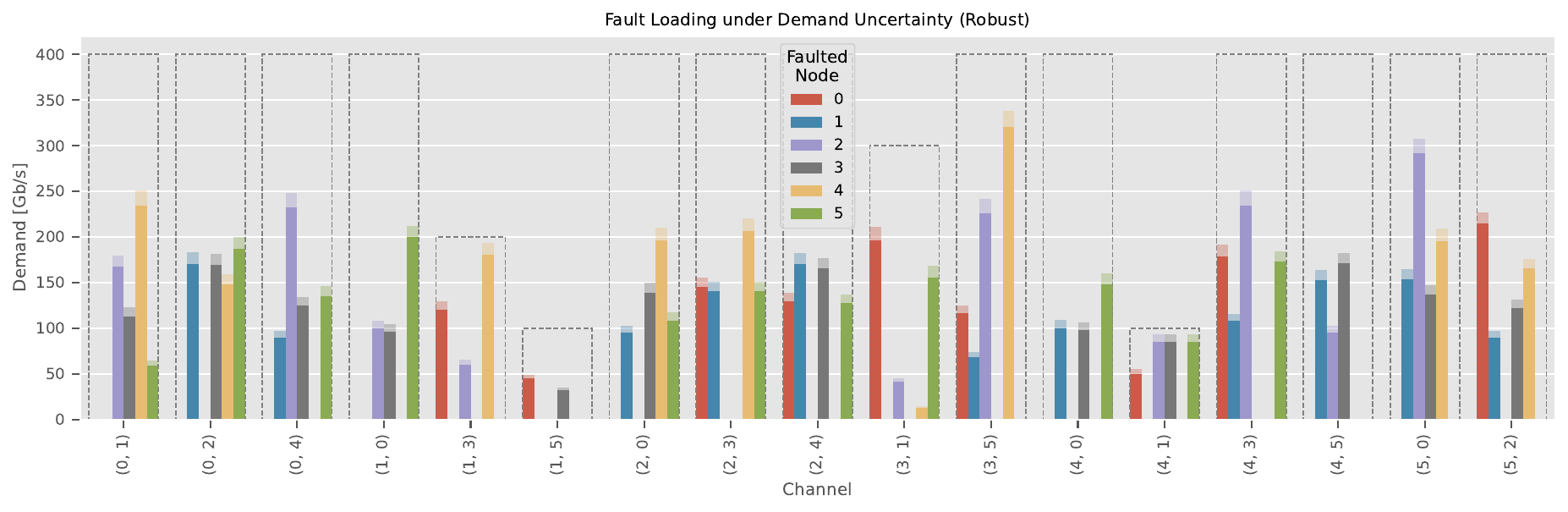}}
\caption{Post-Fault Deviation (Robust)}
\label{robust_dev_post}
\end{figure*}

The results of the post-fault deviation analysis are presented in Fig. \ref{deterministic_dev_post} and Fig. \ref{robust_dev_post}. The deviation analysis was conducted by determining the post-fault loading for each optical channel, given each node failure scenario. After determining the post-fault loading, demand deviation was simulated where each demand assumes a value along the range $[\bar{B_d} - \hat{B_d}, \bar{B_d} + \hat{B_d}]$, where $\bar{B_d}$ denotes the nominal value of demand $d$ and $\hat{B_d}$ denotes the deviation amount of demand $d$. As mentioned in the experiment setup, the parameters selected for implementation yield a robust solution protecting against the worst case of 10 demands deviating by 10\% off their nominal value. To this end, for the demand deviation range discussed above, each demand’s deviation value is within the range $\pm 10\%$ of the nominal value. This deviation is simulated 1000 times, and utilization statistics across all trials are calculated. The objective of these utilization statistics is to determine the general effect parameter deviation can have on the solution and the level to which the robust solution can protect the capacity constraint against said deviation. In the case of Fig. \ref{deterministic_dev_post} and Fig. \ref{robust_dev_post}, the average optical channel loading across the 1000 deviation trials is represented by the solid colour bar, and the maximum loading observed across the 1000 deviation trials is represented by the transparent colour bar.\par
The results presented in Fig. \ref{deterministic_dev_post} consider the post-fault deviation analysis of the deterministic solution. In the previous section, optical channels (1,3), (1,2), and (2,1) of the deterministic were identified as approaching the capacity of the optical channel. These channels specifically are of interest in the deviation analysis. As seen in Fig. \ref{deterministic_dev_post}, when performing the deviation analysis, the maximum loading experienced across these channels exceeds the capacity of these channels, as indicated by the dotted bars. In practice, this translates to a portion of the demands resulting in the overcapacity being blocked, leading to service interruptions.\par
Conversely, Fig. \ref{robust_dev_post} presents the results of the deviation analysis for the robust solution. This analysis shows that even after simulating parameter deviation, none of the observed optical channel loadings exceed the optical channel capacity. In the case of optical channels (4,1) and (1,3), the maximum deviation loading approaches the optical channel capacity; however, there is still a comfortable buffer ensuring the capacity of the optical channel is not exceeded. The results in this section demonstrate how the robust solution protects the optical channel capacity constraint in the face of demand uncertainty.\par

\section{Conclusion}
The work presented in this paper considers a probabilistic and robust fault-tolerant traffic grooming solution for OTN-over-DWDM networks. The developed model is parameterized to ensure the network operator can control the level of resiliency and conservativeness of the solution while simultaneously having the capability to control the amount of reserve capacity to ensure QoS and SLA guarantees are met. Additionally, robust optimization protects the proposed solution against demand uncertainty and, therefore, ensures service continuity during a fault and demand uncertainty scenario. Through the conducted experiment, the flexibility of the proposed robust model is highlighted in various capacities and is directly compared to the deterministic solution. Future work in this field will continue developing management and orchestration practices to support next-generation networks, services and applications.

\section*{Acknowledgement}
The authors would like to acknowledge Mahdi Hemmati and Yuren You from the Huawei Technologies Canada Research Center for their technical support and guidance.
\bibliographystyle{IEEEtran}
\bibliography{sample}

\end{document}